\documentstyle[preprint,aps,version2]{revtex}
\def\baselinestretch{1.5}
\def\half{\frac{1}{2}}

\def\be{\begin{equation}}
\def\ee{\end{equation}}
\def\pa#1{\partial_{#1}}
\def\rq#1{(\ref{eq#1})}
\def\to{\rightarrow}
\def\grad{\nabla}
\newcommand{\spone}{0.9}

\newcommand{\singlespace}{\edef\baselinestretch{\spone}\Large\normalsize}

\newcommand{\conj}{\mbox{c.c.}}

\newcount\sectionnumber
\def\sect
{\global\equationnumber=0
\global\advance\sectionnumber by 1
\the\sectionnumber . }
\newcount\equationnumber
\def   \num
{\eqno{\global\advance\equationnumber by 1
\left(\the\sectionnumber .\the\equationnumber \right)}}

\newcommand{\pn}[4]
 {
   {\renewcommand{\arraystretch}{0.3}
    \begin{array}[b]{@{}c@{}}
      {\scriptstyle {#2}}\\{#1}
    \end{array}
   }
   {\renewcommand{\arraystretch}{0.45}
    \begin{array}[c]{@{}l@{}}
      {\scriptstyle {#3}}\\{\scriptstyle {#4}}
    \end{array}
   }
 }

\newcommand{\pnn}[2]
 {
   {\renewcommand{\arraystretch}{0.3}
    \begin{array}[b]{@{}c@{}}
      {\scriptstyle {#2}}\\{#1}
    \end{array}
   }
 }

\begin{document}

\draft
\begin{title}
\begin{instit}
        Department of Physics
        University of Waterloo
        Waterloo, Ontario
        CANADA N2L 3G1
\end{instit}
\begin{center}\today\end{center}
\begin{center}{WATPHYS TH-92/02}\end{center}
\end{title}
\begin{abstract}
\end{abstract}
\pacs{04.20.Jb, 04.50.+h, 11.17.+y}
\widetext

\section{Introduction}
The study of theories of gravity in two spacetime dimensions can provide
insight into issues in semiclassical and quantum gravity, as these theories
are mathematically much simpler than $(3+1)$ dimensional general relativity
\cite{MST,Arnold,semi,SharTomRobb,MSW,EDW,mannGRG}. Recently two such
theories, that of \cite{MST,Arnold}, referred to as the ``R=T'' theory, and
the string-inspired theory of \cite{MSW,EDW}, have attracted some interest,
due primarily to the fact that their field equations admit
black hole solutions, making them an interesting arena for the study of
quantum gravitational effects.

The latter theory of gravity arises from a non-critical string
theory in two dimensions. Setting to zero the one-loop beta function of the
bosonic sigma model with two target spacetime dimensions gives the effective
target space action
\be
S=\int d^2xe^{-2\Phi}\sqrt{-g}(R - 4(\nabla\Phi)^2 + c). \label{eq2}
\ee
The resultant field equations give rise
to a black hole solution asymmetric about the origin,
\be
ds^2 = -(1-a e^{-Qx})dt^2 + \frac{dx^2}{1-a e^{-Qx}}, \label{eq3}
\ee
with the dilaton field
\be
\Phi  = -\frac{Q}{2}x,   \label{eq4}
\ee
where $Q^2=c$. We have argued \cite{symbh} that from a gravitational
point of view, the asymmetry of \rq3 about the origin is somewhat
objectionable, as it is difficult to see how such a solution could arise
from gravitational collapse of clumped matter (for an alternative
viewpoint see \cite{Callan}).

Matter terms may be incorporated into the action as follows
\be
S=\int d^2x \sqrt{-g}\left\{ e^{-2\Phi}(R-4(\nabla \Phi)^2+{J})
+{\cal L}_M \right\} ,      \label{eq5}
\ee
where ${\cal L}_M$ is a matter Lagrangian,  and ${J}$ is a source term
for the dilaton field.  The above action is a general combination of two
approaches \cite{Callan,Frolov} to coupling matter to the string-inspired
action studied in \cite{MSW,EDW}. From (\ref{eq5}) the field equations are
\be
e^{-2\Phi}(R_{\mu\nu} + 2 \grad_{\mu} \grad_{\nu} \Phi ) = 8\pi G T_{\mu\nu},
\label{eq6}
\ee
\be
R - 4(\grad \Phi )^2 + 4 \grad^2 \Phi + {J} = 0.   \label{eq7}
\ee
The stress-energy tensor $T_{\mu\nu}$ introduced by this procedure may be
regarded as modelling some unknown higher-order effects in the string
theory. It may also be shown that the action \rq5 is equivalent to that of
a massive scalar field $\psi = e^{-\Phi}$ non-minimally coupled to
curvature \cite{symbh}, which allows a less ambiguous interpretation of the
matter term. If an appropriate point source of matter at the origin is
introduced, the solution of (\ref{eq6},\ref{eq7}) is a symmetric version of
the above black hole solution, that is (\ref{eq3},\ref{eq4}) with $x$
replaced by $|x|$. We have shown that this solution will result from a
collapsing dust if we include appropriate surface stresses and dilaton
charges \cite{symbh}. The latter may be generated by the source ${J}$ for
this dilaton charge; in the dilaton vacuum $J=c$. The properties of this
black hole solution have been studied in detail in refs.
\cite{symbh,match}; they are broadly similar to those found before for the
black hole solution of the R=T theory, although there are significant
differences.

We regard the theory described by the action (\ref{eq5}) as an interesting
theory of two-dimensional gravitation in its own right.
The purpose of the present paper is to explore the dynamical properties of
this theory in more detail, comparing them to the similar treatment of
the R=T theory in ref. \cite{Arnold}. We find that the theory has a
sensible post-Newtonian expansion. Consideration of the weak field limit
shows that there will be gravitational radiation, as the trace of the
perturbation of the metric obeys a field equation with source terms from
the stress-energy and the dilaton source $J$.
The equation for stellar equilibrium in this system is
obtained. Finally, we consider the cosmological solutions of the field
equations, and find that while a dust-filled universe will eventually
collapse, the radiation-filled universe cannot collapse at any time. We
summarize our results and discuss further areas of interest in a concluding
section.

\section{Post-Newtonian Calculations}

We wish to demonstrate that this theory has a sensible Newtonian and
post-Newtonian limit. We see that in the first approximation Newton's theory
holds, and the higher-order terms are qualitatively similar to those found
before for the R=T theory \cite{Arnold}.

We consider a system of particles experiencing mutual gravitational
attraction, and let $\bar{M}, \bar{r}$ and $\bar{v}$ be typical values of
their masses, separations and speeds. Comparison with Newton's theory of
gravity in one spatial dimension yields the Newtonian potential
\be
\xi = 2\pi G\bar{M}|x|,   \label{eq8}
\ee
where a constant of integration has been ignored. If we consider a
test particle falling into this potential from $|x|=\bar{r}$ initially at rest,
its maximum speed will be
\be
\bar{v}^2 \sim G\bar{M}\bar{r},    \label{eq9}
\ee
which gives an approximate relation between $\bar{M}, \bar{r}$ and
$\bar{v}$ for a system of particles. The Newtonian approximation gives the
first-order terms in the small parameter $\bar{v}^2$, so the objective of
the post-Newtonian approximation is to supply the higher order terms in the
expansion of physical quantities.

We expect the metric to be approximately Minkowskian where gravity is
weak, but we do not
assume that it has any particular form. The expansion of the metric is
\begin{eqnarray}
g_{00}&=&-1+\pn{g}{2}{}{00}+\pn{g}{4}{}{00}+\cdots ; \\
g_{01}&=&\pn{g}{3}{}{01}+\pn{g}{5}{}{01}+\cdots ;\\
g_{11}&=& 1+\pn{g}{2}{}{00}+\pn{g}{2}{}{00}+
              \pn{g}{4}{}{00}+\cdots ,
\label{eq10a}
\end{eqnarray}
where $\pn{g}{N}{}{\ \ \mu\nu}$ denotes the term of order $\bar{v}^N$ in the
expansion. We can then calculate the Christoffel symbols and thus the
required components of the Ricci tensor,
\be
\pn{R}{2}{}{00}= -\half \pa1\pa1 \pn{g}{2}{}{00},    \label{eq13}
\ee
\be
\pn{R}{4}{}{00} = -\half [ \pa1\pa1 \pn{g}{4}{}{00} -\half\pa1 \pn{g}{2}{}{00}
\pa1 \pn{g}{2}{}{11} -\pa0\pa0
\pn{g}{2}{}{00} + \pa1 \pn{g}{2}{}{00} \pa1 \pn{g}{2}{}{00} ],    \label{eq14}
\ee
\be
\pn{R}{3}{}{01} = -\half \pa1 \pa1 \pn{g}{3}{}{01},     \label{eq15}
\ee
\be
\pn{R}{2}{}{11} = -\half \pa1 \pa1 \pn{g}{2}{}{11},    \label{eq16}
\ee
where $\pn{R}{N}{}{\ \ \mu\nu}$ denotes the term of order $\bar{v}^N/\bar{r}^2$
in the expansion of $R_{\mu\nu}$. We interpret $T^{00}$, $T^{01}$ and
$T^{11}$ as the energy density, momentum density and momentum flux, which
leads us to make the following expansions:
\be
T_{00} = \pn{T}{0}{}{00} + \pn{T}{2}{}{00} + \cdots,  \label{eq17}
\ee
\be
T_{01} = \pn{T}{1}{}{01} + \pn{T}{3}{}{01}+\cdots,  \label{eq18}
\ee
\be
T_{11} = \pn{T}{2}{}{11} + \pn{T}{4}{}{11} +\cdots.   \label{eq19}
\ee
{}From our experience with the black hole solutions, we also expect
\be
\Phi = \pnn{\Phi}{2}+\pnn{\Phi}{4}+\cdots,    \label{eq20}
\ee
\be
J = c+\pnn{J}{2}+\pnn{J}{4}+\cdots, \label{eq21}
\ee
where $\Phi$ is the dilaton field and $J$ is the dilaton current.

The field equations can be expanded in powers of our small parameter, giving
us the forms
\be
\pn{R}{N}{}{\mu\nu} + 2\pnn{(\nabla_\mu\nabla_\nu \Phi)}{N}
= 8\pi G \pn{T}{(N-2)}{}{\mu\nu},
\label{eq22}
\ee
\be
\pnn{R}{N} - 4 \pnn{[(\nabla \Phi)^2]}{N} + 4 \pnn{(\nabla^2 \Phi)}{N}
+ \pnn{J}{N}=0, \label{eq23}
\ee
where $N$ is the order in $\bar{v}^2/\bar{r}$. For the Newtonian
approximation, we only need to determine $\pn{g}{2}{}{\ \ 00}$, so we will only
need the $\mu=0,\nu=0$ component of \rq{22} to order $N=2$. For the post-
Newtonian expansion, we will need the 00 component of \rq{22} to order 4,
the 01 component to order 3, and the 11 component to order 2, as well as
\rq{23} to order 2.

First, we compute the Newtonian approximation. The 00 component of \rq{22} to
order 2 gives
\be
\pa1\pa1 \pn{g}{2}{}{00} = -16\pi G \pn{T}{0}{}{00},  \label{eq24}
\ee
which has solution $\pn{g}{2}{}{\ \ 00} = -4\xi$, where
\be
\xi(x,t) = 2\pi G \int dx' \pn{T}{0}{}{00} (x',t) |x-x'|      \label{eq25}
\ee
is the Newtonian potential. Note that this differs from the result
$\pn{g}{2}{}{\ \ 00} = -2\xi$ of \cite{Arnold} by a factor of two.

We now compute the post-Newtonian terms, $\pn{g}{2}{}{\ \ 11},\:
\pn{g}{3}{}{\ \ 01},\: \pn{g}{4}{}{\ \ 00}$
and $\pnn{\Phi}{2}$. The 11 component of \rq{22} to order 2 gives
\be
\pa1\pa1 \pn{g}{2}{}{11} = 4 \pa1\pa1 \pnn{\Phi}{2},  \label{eq26}
\ee
which has solution $\pn{g}{2}{}{\ \ 11} = 4 \pnn{\Phi}{2}$. If we now consider
the
dilaton equation \rq{23} to order 2 we get
\be
\half \pa1 \pa1 \pn{g}{2}{}{00} -\half \pa1\pa1 \pn{g}{2}{}{11}+ 4\pa1\pa1
\pnn{\Phi}{2} +\pnn{J}{2}=0,
\label{eq27}
\ee
and substituting $\pn{g}{2}{}{\ \ 11} = 4 \pnn{\Phi}{2}$ gives
\be
\pa1\pa1 \pn{g}{2}{}{11} = -\pa1\pa1 \pn{g}{2}{}{00} - \pnn{J}{2}  = 4\xi'' -
\pnn{J}{2},   \label{eq28}
\ee
so the solution is $\pn{g}{2}{}{\ \ 11} = 4\beta$, and thus $\pnn{\Phi}{2} =
\beta$, where $\beta$ is a new field defined by
\be
\beta (x,t) = \int dx' |x-x'| ( \half \xi'' - \frac{1}{8} \pnn{J}{2}).
\label{eq29}
\ee
Note in particular that if $\pnn{J}{2}=0$, $\beta = \xi$.

We now take the 01 component of \rq{22} to order 3 to determine
$\pn{g}{3}{}{\ \ 01}$. This gives
\be
\pa1\pa1 \pn{g}{3}{}{01}  = -16\pi G \pn{T}{1}{}{01} + 4\pa1\pa0 \pnn{\Phi}{2}
= -16\pi G \pn{T}{1}{}{01} +
4\pa0\pa1\beta, \label{eq30}
\ee
which gives $\pn{g}{3}{}{\ \ 01}= \eta$, where $\eta$ is a new field defined by
\be
\eta(x,t) = \int dx' |x-x'| (-8\pi G \pn{T}{1}{}{01}(x',t) + 2\dot{\beta}').
\label{eq31}
\ee

Finally, we compute $\pn{g}{4}{}{\ \ 00}$ from the 00 component of \rq{22} to
order 4,
substituting for $\pn{g}{2}{}{\ \ 11}$ and $\pnn{\Phi}{2}\ $ from above:
\be
\pa1\pa1 \pn{g}{4}{}{00}
=-16\pi G\pn{T}{2}{}{00}+4(\ddot{\beta}-\ddot{\xi})-16\xi'^2,
\label{eq32}
\ee
which gives $\pn{g}{4}{}{\ \ 00} = \psi$, if we define the new field $\psi$ by
\be
\psi = \int dx' |x-x'|
(2(\ddot{\beta}-\ddot{\xi}) -8\xi'^2- 8 \pi G \pn{T}{2}{}{00}).
\label{eq33}
\ee
It is also perhaps worth noting that
\be
\ddot{\beta}-\ddot{\xi} =
-\frac{1}{8} \int dx' |x-x'| \frac{\partial^2\pnn{J}{2}}{\partial t^2},
\label{eq34}
\ee
which will vanish if $\pnn{J}{2}\ $ is linear in time.

This completes the calculation of the post-Newtonian approximation.
It should also be noted that because this is a
truncated series in powers of the distance $r$, it will be a better
approximation near the system, even though we expect the spacetime to be
asymptotically flat for constant $J$ and $T_{\mu\nu}=0$ \cite{symbh}.
The main
relevance of this calculation is that it shows that one can carry out
an expansion in this theory about its Newtonian limit. Expansion at large
distances ({\it i.e.} about the asymptotically flat solution) was
considered in \cite{EDW}; we shall not pursue this issue any further here.

\section{Weak Field Approximation}

We now turn to the weak-field approximation, and demonstrate the existence of
gravitational radiation in the linear fields. We consider the metric to
be a perturbation on a Minkowski background,
\be
g_{\mu\nu} = \eta_{\mu\nu} + h_{\mu\nu}.    \label{eq35}
\ee
If we write the field equation as the trace and traceless parts, the
equations are
\be
R^{(1)} + 2\nabla^2 \Phi = 8 \pi G T,    \label{eq36}
\ee
\be
2\nabla_\mu\nabla_\nu \Phi -g_{\mu\nu}\nabla^2 \Phi = 8\pi G ( T_{\mu\nu}
-\half g_{\mu\nu} T),    \label{eq37}
\ee
\be
R^{(1)} + 4\nabla^2 \Phi + J = 0,   \label{eq38}
\ee
where $R^{(1)}$ is the linear order Ricci scalar, and all the non-linear terms
have been neglected. We can rewrite these equations as:
\be
R^{(1)}  = 16\pi GT + J, \label{eq39}
\ee
\be
2\nabla_\mu\nabla_\nu\Phi =8\pi G(T_{\mu\nu}-g_{\mu\nu}T) - \half
g_{\mu\nu}J. \label{eq40}
\ee

Imposing the coordinate condition
\be
\pa{\mu}\pa{\nu}h^{\mu\nu} = \half\pa{}^2 h, \label{eq41}
\ee
which was used in \cite{Arnold}, \rq{39} becomes
\be
\pa{}^2 h = -32 \pi GT - 2J,   \label{eq42}
\ee
which is a wave equation with source term in the trace $h = h_\mu^\mu$ of
the metric perturbation.   Note that this differs from the analogous
equation in \cite{Arnold} by the ubiquitous factor of two, and by the fact
that the dilaton source may act as a source of gravitational radiation in
the absence of matter.  The particular integral solutions of this equation
are given in terms of retarded and advanced potentials
\be
h(x,t)=\pm 16\pi G\int dx' \int^t dt' \:
       \left[ {\cal F}(x',t'\mp|x-x'|) \right],
       \label{potls}
\ee
to which any solution of the corresponding homogeneous system ((\ref{eq42})
with $T=0=J$) may be added. Here ${\cal F}(x,t)=T(x,t)+\frac{1}
{16\pi G}J(x,t)$.

The concept of a gravitational wave in the vacuum is somewhat problematic,
in part because the notion of a vacuum depends upon whether or not one
considers the vacuum to be that region of spacetime for which
$T_{\mu\nu}=0$ and $J=0$ or for which $T_{\mu\nu}=0$ and $J=c$.

In the former case spacetime is flat outside of matter.
Consider a system of
oscillating matter which has an energy-momentum tensor trace  and dilaton
current representable as a Fourier integral or sum over frequencies
$\omega$. A single Fourier component is
\be
{\cal F}(x,t)={\cal F}(x,\omega)\exp(-i\omega t)+\conj ,
\ee
and the retarded potential solution (\ref{potls}) giving gravitational
waves is then
\be
h(x,t)=8\pi G\int dx' \int^t dt' \,
{\cal F}(x',\omega)\exp\left[-i\omega(t'-|x-x'|)\right]+\conj \ ,
\label{gwave}
\ee
where ``$\conj$'' denotes the complex conjugate
of the preceding term.
If the source is finite, with maximum extension $R=|x'|$, and we are
situated in a region of space outside of the source with
$r\equiv|x|>R$,  then the gravitational wave
is a plane wave
\be
h=H \exp(i k_\mu x^\mu) + \conj , \label{planewave}
\ee
with amplitude and wave vector
\begin{eqnarray}
H&=& 8\pi G i \omega^{-1} \int dx' \,
{\cal F}(x',\omega)\exp(-i k_1 x') \nonumber \\
&\equiv&8\pi Gi\omega^{-1}\,{\cal F}(k_1,\omega) \label{amplitude}; \\
k_0&=&-\omega ;\;\;\; k_1=\omega\hat{x}\label{wavevector} ,
\end{eqnarray}
where the complete Fourier transform of the energy-momentum tensor trace is
defined. Here we have used $|x-x'|=r-x'\hat{x}$ with $\hat{x}\equiv x/r$.

Here, as for the theory considered in \cite{Arnold},
it is the global nature of the gravitational wave which contains
the non-trivial physics. Such waves are co-ordinate waves locally.
Although on either side of the source one may choose a coordinate
system which ``travels with the wave'', outside of the
source the wave is of the form
\be
h=H \exp [-iw(t-|x|)],
\ee
and so such a coordinate transformation cannot be applied globally.
An observer crossing the source will see a flip in the wave's direction
of propagation. In this sense there is gravitational
radiation in the weak field approximation.

For the case $T_{\mu\nu}=0$ and $J=c$ (the `dilaton vacuum'), the full
system of equations yields in general the unique solution
(\ref{eq3},\ref{eq4}).  The solution to (\ref{eq42}) in this case is
\be
h=-\frac{c}{2}(x^2-t^2) + f_1(x-t) + f_2(x+t) \label{stringweak}
\ee
which is the weak field approximation to the exact solution
\be
\sigma=\frac{M}{\sqrt{c}}-\frac{c}{2}(x^2-t^2)  \label{stringexact}
\ee
of (\ref{eq6},\ref{eq7})
where sigma is the conformal factor for the metric
$ds^2=\sigma(-dt^2+dx^2)$. Here $M/\sqrt{c}$ may be interpreted as the mass
parameter of the black hole \cite{EDW,symbh}, as follows from a computation
of the ADM mass. The solution (\ref{stringexact}) is unique up to
coordinate transformations, and may be transformed to the solution
(\ref{eq3}) under a suitable change of coordinates.
For small $M/\sqrt{c}$, the solution (\ref{stringweak}) is recovered. In
this case we do not have wavelike solutions appearing, due to the presence
of the constant dilaton source $c$.

In both cases the other field equation \rq{40} will give three further
constraints on the linear dilaton field $\Phi$ and the components of the
metric perturbation, thus completely determining the solution.  The energy
and momentum carried by the wave  cannot be computed without solving this
system. We could attempt to construct a second-order weak-field system, but
this is too complicated to be an enlightening exercise, so we
shall not consider this problem any further.

\section{Stellar Structure}

We now consider the equations of fluid equilibrium, which govern the
existence of `stars' (clumps of matter in one spatial dimension)
in the theory. We use the static metric of the form
\be
ds^2 = -B^2(x) dt^2 + dx^2  \label{eq43}
\ee
together with the field equations
\be
R + 2\nabla^2\Phi = 8\pi GT e^{2\Phi},   \label{eq44}
\ee
\be
2\nabla_\mu\nabla_\nu\Phi -g_{\mu\nu}\nabla^2\Phi=8\pi G
(T_{\mu\nu}-\half g_{\mu\nu}T) e^{2\Phi},    \label{eq45}
\ee
\be
R-4(\nabla\Phi)^2 + 4\nabla^2 \Phi +J =0.   \label{eq46}
\ee
The metric gives
\be
R=-2\frac{B''}{B}\mbox{ and }\nabla^2\Phi =\Phi'' +\frac{B'}{B}\Phi'
\label{eq47}
\ee
if the stress-energy is the perfect fluid energy-momentum tensor. The field
equation \rq{45} gives
\be
\Phi''-\frac{B'}{B}\Phi' = 4\pi G (p+\rho) e^{2\Phi},   \label{eq48}
\ee
where $p$ is the pressure, $\rho$ is the density, and $'=d/dx$. Also,
the field equations \rq{44} and \rq{46} give
\be
\Phi'^2 = 4\pi G(p-\rho) e^{2\Phi} +\frac{J}{4} +\half\frac{B''}{B},
\label{eq49}
\ee
so we may obtain an expression for $\nabla^2\Phi$. Substitution of this
expression in \rq{44} gives an equation quadratic in $B''/B$, whose roots are
\be
\frac{B''}{B} = 8 \pi G \rho e^{2\Phi} + \left(\frac{p'}{p+\rho}\right)^2 \pm
\left(\frac{p'}{p+\rho}\right) \left[\left(\frac{p'}{p+\rho}\right)^2 +
16\pi Gpe^{2\Phi}+J \right]^{1/2}\!\!\!\!\!\!.  \label{eq50}
\ee

We also have the equation of hydrostatic equilibrium,
\be
-p' = (p+\rho)\left[\ln(-g_{00})^{1/2}\right]', \label{eq51}
\ee
which in our case reduces to
\be
\frac{B'}{B} = -\frac{p'}{p+\rho}.  \label{eq52}
\ee
Thus the general equation for hydrostatic equilibrium in our system is
\be
-p'' = (p+\rho)\frac{B''}{B} - \frac{p'(2p'+\rho')}{p+\rho}, \label{eq53}
\ee
where $\frac{B''}{B}$ is given by \rq{50}. Given an equation of state
$p=p(\rho)$, the solution of \rq{53} will give $\rho=\rho(r)$ and hence the
metric.

If we compare this to Newton's equation of stellar equilibrium
\be
-p'' = 4\pi G\rho^2 - \frac{p'\rho'}{\rho},      \label{eq54}
\ee
we see that in the Newtonian limit $p \to 0$, $J \to 0$, $\Phi \to 0$
\rq{50} reduces to
\be
\frac{B''}{B} = 8 \pi G \rho.    \label{eq55}
\ee
Thus in this limit our equation \rq{53} will only differ from the Newtonian
equation \rq{54} by a factor of two multiplying $G$.

\section{Cosmology}

Consider the two-dimensional Robertson-Walker metric
\be
ds^2 = -dt^2 + a^2(t)\frac{dx^2}{1-k x^2},  \label{eq56}
\ee
with the field equations (\ref{eq44},\ref{eq45},\ref{eq46}). As previously
noted in ref. \cite{Arnold}, in two dimensions we do not have three different
cosmological models corresponding to open, closed and flat spacetimes since
the denominator in the second term in (\ref{eq56}) can be absorbed into a
definition of the spatial coordinate.

If we assume the perfect fluid stress-energy, \rq{44} and \rq{45} become
\be
e^{-2\Phi} ( -\ddot{\Phi} + \frac{\dot{a}}{a}\dot{\Phi}) = -4 \pi G(p+\rho),
\label{eq57}
\ee
\be
e^{-2\Phi}(2\ddot{a}-2\dot{\Phi}\dot{a}-2\ddot{\Phi}a) = 8 \pi G(p-\rho)a.
\label{eq58}
\ee
If we now take the equation of state to be $p=(\gamma-1)\rho$, then
conservation of energy $T^{\mu\nu}_{;\nu}=0$ will give $a\rho^\gamma =
a_0\rho_0^\gamma$, and substitution in (\ref{eq57},\ref{eq58}) will give
\be
\ddot{a} = \frac{2}{\gamma} \dot{a}\dot{\Phi}
+2 \frac{\gamma-1}{\gamma} a\ddot{\Phi}, \label{eq59}
\ee
and
\be
e^{-2\Phi} (-\frac{\ddot{a}}{a}+2\ddot{\Phi}) = 8\pi G\rho .    \label{eq60}
\ee
We also find that \rq{46} becomes
\be
\ddot{a}+2a\dot{\Phi}^2-2a\ddot{\Phi}-2\dot{a}\dot{\Phi}+\half Ja = 0.
\label{eq61}
\ee
Solution of these equations for general $\gamma$ is quite difficult.
However a solution may easily be obtained for two special cases, $\gamma=1$
and $\gamma=2$, corresponding to dust- and radiation-filled spacetimes
respectively. For simplicity we shall henceforth consider only $J=c$
solutions.

The case $\gamma=1$ was studied in \cite{symbh}, but we include the results
here in the interests of completeness.
We find \rq{59} becomes
\be
\ddot{a} = 2\dot{a}\dot{\Phi},   \label{eq62}
\ee
and substitution of this in \rq{61} gives
\be
\ddot{\Phi} - \dot{\Phi}^2 = \frac{c}{4},   \label{eq63}
\ee
with solution
\be
\Phi (t) = -\ln \left(\cos(\frac{Q}{2}t+\beta)\right) + \Phi_0,  \label{eq64}
\ee
where $Q^2=c$, and $\beta$ and $\Phi_0$ are constants of integration. If we
substitute this solution into \rq{62}, we find that $a$ is given by
\be
a(t)= -\lambda \tan(\frac{Q}{2}t+\beta) + \alpha,  \label{eq65}
\ee
where $\lambda$ and $\alpha$ are constants, and thus \rq{60} implies $\rho_0$
is
given by
\be
\rho_0 = \frac{Q^2 \alpha}{16 \pi G} e^{-2\Phi_0}.    \label{eq66}
\ee
Note that we chose $\alpha=1$ and $\beta=0$ for convenience in the previous
paper
\cite{symbh}. We see that this solution
will collapse at
\be
t_c = \frac{2}{Q} \tan^{-1}(\frac{\alpha}{\lambda})-\beta,  \label{eq67}
\ee
and as the density varies inversely with $a$, the density will diverge at the
collapse. The curvature
\be
R = - \frac{Q^2\lambda \sec^2(\frac{Q}{2}t+\beta)\tan^2(\frac{Q}{2}t+\beta)}
{\alpha -\lambda \tan(\frac{Q}{2}t+\beta)}      \label{eq68}
\ee
also diverges at $t_c$. This collapse is similar that of the cosmological
solution of the R=T theory \cite{Arnold}. However, in contrast to
\cite{Arnold}, one cannot construct a two-dimensional analogue of the FRW
cosmology in general relativity using \rq{65}, since $a$ vanishes at only one
time $t_c$, and diverges at finite times both before and after $t_c$.

In the case $\gamma=2$ (radiation), \rq{59} becomes
\be
\ddot{a} = \dot{a}\dot{\Phi} + a\ddot{\Phi}, \label{eq69}
\ee
which may be integrated to give
\be
\dot{a}=a\dot{\Phi}+B,  \label{eq70}
\ee
$B$ a constant of integration. If we substitute this in \rq{61}, we may obtain
\be
\ddot{a} = \frac{2(\dot{a}-B)^2}{a} + \half ca. \label{eq71}
\ee

Considering first solutions with $B=0$, we find
\be
\Phi(t) = -\ln \left(\cos(\frac{Q}{\sqrt{2}}t+\beta)\right) + \Phi_0,
\label{eq72}
\ee
\be
a(t) = A \sec( \frac{Q}{\sqrt{2}}t+\beta),  \label{eq73}
\ee
where $\Phi_0$, $\beta$ and $A$ are arbitrary constants. This solution
obviously does not collapse, and corresponds to a universe which is
expanding.

Turning now to the more general case, we
may obtain a first integral of \rq{71} by treating it as a differential
equation for $\dot{a}$ as a function of $a$. The implicit solution is
\be
\ln(ca^2+4B^2-4By)+\frac{4By}{ca^2+4B^2-4By}=\ln(a^2) + \kappa,
\label{eq74}
\ee
where $y=\dot{a}(a)$, and $\kappa$ is a constant. We can rewrite this in a
simpler form by scaling $\hat{a}=\sqrt{c}a/{2B},  \hat{y} = y/B$,
which gives
\be
\ln(\hat{a}^2+1-\hat{y})+\frac{\hat{y}}{\hat{a}^2+1-\hat{y}}=\ln(\hat{a}^2)+
k, \label{eq75}
\ee
$k$ a constant.

We check now for collapsing solutions.
If the solution given implicitly by \rq{75} collapses, it follows that this
equation must be satisfied as $a \to 0$. Thus, we let
$\hat{a}^2=\epsilon$, $\hat{y}=1+\epsilon -\delta$, and consider \rq{75} as
$\epsilon \to 0$. We find \rq{75} may be rewritten as
\be
\ln(\delta)+\frac{1}{\delta} = \ln(\epsilon) + k,  \label{eq76}
\ee
which implies
\be
-\frac{1}{\epsilon}e^{-k}=-\frac{1}{\delta}e^{-1/\delta} \geq
-\frac{1}{e}, \label{eq77}
\ee
since $xe^x \geq -1/e$ for all $x$. Thus
\be
\epsilon e^{k} \geq e, \label{eq78}
\ee
which cannot be satisfied for finite $k$ as $\epsilon \to 0$. Thus,
\rq{75} cannot be satisfied as $a \to 0$, so the solution
described by \rq{75} cannot collapse.

Alternatively, since \rq{71} implies that $\ddot{a}$ is
always positive, collapse is impossible if $\dot{a}$ is initially
positive. If $\dot{a}$ is initially negative but vanishes before the collapse
occurs (i.e., for $a \in (0,1]$), collapse is also impossible. Thus, if the
initial condition is $\hat{y}( \hat{a}^2=1) = y_0$, then
\be
k = \ln(2+y_0)+\frac{2}{2+y_0} -1, \label{eq79}
\ee
and collapse is possible only if $\hat{y}$ does not vanish for
$\hat{a}^2 \in (0,1]$ (note that $\hat{y}$ and $\dot{a}$ do not
necessarily have the same sign). When $\hat{y}=0$, $\hat{a}$ is given by
\be
\hat{a}^2 = \frac{1}{e^{k}-1}, \label{eq80}
\ee
and thus $\hat{y}=0$ for $\hat{a}^2 \in (0,1]$ if $e^{k} \geq 2$,
which implies
\be
 - \frac{2}{2+y_0}e^{-2/(2+y_0)} \geq -\frac{1}{e}.  \label{eq81}
\ee
This is true for all $y_0$; therefore collapse is impossible.

Hence although we are unable to obtain the full solution $a(t)$
for a radiation-filled spacetime, we have been able to demonstrate
such a universe in general cannot collapse. Time-reversal invariance
therefore implies that the scale factor reaches a minimal value at some
time $t=t_{\rm min}$.

\section{Conclusions}

The above considerations, when combined with the results in refs.
\cite{symbh,match} on gravitational collapse, indicate that the
two-dimensional theory of gravity given by the action \rq5 yields a
classical theory of gravity which is as rich in structure as the R=T theory
proposed earlier \cite{Arnold}.
Of course from a classical relativist's viewpoint, the presence of the
dilaton introduces features which are markedly different from the `R=T'
theory.  In the latter case, curvature is generated solely by stress-energy
which is prescribed from the matter Lagrangian. In the
string-motivated theory studied here, the dilaton field cannot be decoupled
from the remaining gravity-matter system: even a vanishing dilaton field
imposes constraints on the stress energy in addition to those which follow
from the conservation laws. Indeed, upon reparametrizing the dilaton field
so that $e^{-2\Phi}=\varphi$, it is easily seen that the action
(\ref{eq5}) (with $J=0$) is two dimensional Brans-Dicke theory with
$\omega=-1$.

The post-Newtonian  expansion is similar to
that of general relativity, although the presence of the dilaton field
introduces some extra complications. In the weak-field approximation the
trace of the metric perturbation obeys a wave equation, demonstrating the
existence of gravitational `radiation' in this theory. The equation of
stellar equilibrium was obtained, and we saw that in the Newtonian limit,
it reduced to a form similar to the Newtonian equation. The Newtonian
approximations to this theory were identical to those of \cite{Arnold}
apart from a factor of two multiplying $G$.

The development of the cosmological solution in \cite{symbh} was included.
This study shows a dust-filled spacetime will collapse in finite proper
time for suitable initial conditions, with divergent density and curvature,
although  a solution which collapses cannot have developed from an initial
singularity, and vice-versa. This solution is the basis for the
demonstration that the symmetric black hole solution arises from
gravitational collapse \cite{symbh,match}. In the case of a radiation-filled
spacetime,  we were unable to solve the equations in general,
although we were able to show that the spacetime never collapses. The
collapsing dust solution has similarities to the collapsing dust in the R=T
theory \cite{Arnold}, but the general properties of the cosmological
solutions are quite different.

\section{Acknowledgements}
This work was supported in part by the Natural Sciences and Engineering
Research Council of Canada.

\newpage
\singlespace

\end{document}